\documentclass[a4paper,oneside,10pt]{article}
\usepackage[utf8]{inputenc}
\usepackage[T1]{fontenc}
\usepackage{mathptmx}
\usepackage[english]{babel}
\usepackage[a4paper,left=25mm,right=25mm,top=25mm,bottom=25mm]{geometry}
\usepackage{graphicx}
\usepackage{xcolor}
\usepackage{textcomp}
\usepackage{amsmath,amssymb}
\usepackage{cite}
\usepackage[hidelinks]{hyperref}
\usepackage[font=normalsize,labelfont=normalfont,labelsep=period,justification=centering,singlelinecheck=false]{caption}
\usepackage{acronym}
\acrodef{OLS}{Open Line System}
\acrodef{ROADM}{Reconfigurable Optical Add-Drop Multiplexer}
\acrodef{BVT}{Bandwidth Variable Transceiver}
\acrodef{OSNR}{Optical Signal to Noise Ratio}
\acrodef{ASE}{Amplified Spontaneous Emission}
\acrodef{NETCONF}{Network Configuration}
\acrodef{NLI}{Non Linear Interference}
\acrodef{BER}{Bit Error Rate}
\acrodef{OSaaS}{Optical Spectrum as a Service}
\acrodef{CFO}{Carrier Frequency Offset}
\acrodef{CDC}{Chromatic Dispersion Compensation}
\acrodef{DGD}{Differential Group Delay}
\acrodef{ORP}{Optical Received Power}
\acrodef{SNR}{Signal to Noise Ratio}
\acrodef{PDL}{Polarization Dependent Loss}
\acrodef{ML}{Machine Learning}
\acrodef{ANN}{Artificial Neural Network}
\acrodef{OCM}{Optical Channel Monitor}
\acrodef{OPM}{Optical Performance Monitoring}
\acrodef{EDFA}{Erbium Doped Fiber Amplifier}
\acrodef{EON}{Elastic Optical Network}
\acrodef{DWDM}{Dense Wavelength Division Multiplexing}
\acrodef{QoT}{Quality of Transmission}
\acrodef{WSS}{Wavelength Selective Switch}
\acrodef{OOK}{On-Off Keying}
\acrodef{PSD}{Power Spectral Density}
\acrodef{SLA}{Service Level Agreement}
\acrodef{XPM}{Cross Phase Modulation}
\acrodef{CPM}{Cross Polarization Modulation}
\acrodef{SOP}{State of Polarization}
\acrodef{PD}{Photodetector}
\acrodef{ILA}{In-Line Amplifier}
\acrodef{DAS}{Distributed Acoustic Sensing}
\acrodef{DL}{Deep Learning}
\acrodef{FPN}{Feature Pyramid Network}
\acrodef{MIL}{Multiple-Instance Learning}
\acrodef{BCE}{Binary Cross-Entropy}
\acrodef{ASN}{Alcatel Submarine Networks}
\acrodef{SNR}{Signal-to-Noise Ratio}
\acrodef{AIS}{Automatic Identification System}
\acrodef{EFBL}{Emerald Fibre Bridge Link}
\acrodef{CPA}{Closest Point of Approach}
\acrodef{INTGN}{Irish National Tide Gauge Network}
\acrodef{CDF}{Cumulative Distribution Function}

\makeatletter
\renewcommand{\section}{\@startsection{section}{1}{\z@}%
  {-12pt plus -2pt minus -2pt}{3pt}%
  {\normalfont\normalsize\bfseries}}
\renewcommand{\subsection}{\@startsection{subsection}{2}{\z@}%
  {-6pt plus -1pt minus -1pt}{3pt}%
  {\normalfont\normalsize\bfseries}}
\renewcommand{\@seccntformat}[1]{\csname the#1\endcsname\quad}
\renewcommand{\maketitle}{%
  \begingroup
  \thispagestyle{ictonfirst}%
  \begin{center}
    {\fontsize{16}{19}\selectfont\bfseries\@title\par}
    \vspace{12pt}
    {\normalsize\makebox[\linewidth][c]{\makebox[1.2\textwidth][c]{\@author}}\par}
  \end{center}
  \vspace{6pt}
  \endgroup
}
\def\ps@ictonfirst{%
  \let\@oddhead\@empty
  \let\@evenhead\@empty
  \def\@oddfoot{\scriptsize 979-8-3195-4420-9/26/\$31.00 \textcopyright2026 European Union\hfil}%
  \let\@evenfoot\@oddfoot}
\renewenvironment{thebibliography}[1]
  {\section*{REFERENCES}%
   \list{\@biblabel{\@arabic\c@enumiv}}%
     {\settowidth\labelwidth{\@biblabel{#1}}%
      \leftmargin\labelwidth
      \advance\leftmargin\labelsep
      \usecounter{enumiv}%
      \let\p@enumiv\@empty
      \renewcommand\theenumiv{\@arabic\c@enumiv}%
      \setlength{\itemsep}{0pt}%
      \setlength{\parsep}{0pt}%
      \setlength{\partopsep}{0pt}%
      \setlength{\topsep}{3pt}}%
   \sloppy\clubpenalty4000\widowpenalty4000\sfcode`\.\@m}
  {\endlist}
\makeatother

\newenvironment{icton_abstract}
  {\vspace{6pt}\par\begingroup\setlength{\parindent}{0pt}\noindent\textbf{ABSTRACT}\par\noindent\ignorespaces}
  {\par\endgroup}

\newcommand{\keywords}[1]{%
  \par\noindent\textbf{Keywords}: #1\par}

\begin{document}
\selectlanguage{english}

\title{DAS–AIS Association Patterns for Vessel Monitoring on an Operational Subsea Fibre Link}

\author{
  Tian Tian \textsuperscript{(1)},
  Agastya Raj \textsuperscript{(1)},
  Lara Flanagan \textsuperscript{(1)},
  Nicolas Celli \textsuperscript{(3)},
  Chris Bean \textsuperscript{(3)},
  John Kennedy \textsuperscript{(2)},
  Marco Ruffini~\textsuperscript{(1)}
}

\maketitle
\vspace{-10pt}
\noindent\parbox{\textwidth}{\centering
  \small
  \makebox[\linewidth][c]{\makebox[1.2\textwidth][c]{\textsuperscript{(1)}IRIS Research Group, ADAPT Research Centre, School of Computer Science and Statistics, Trinity College Dublin, Ireland}}\\
  \makebox[\linewidth][c]{\makebox[1.2\textwidth][c]{\textsuperscript{(2)}School of Engineering, ADAPT Research Centre, Trinity College Dublin, Ireland}}\\
  \makebox[\linewidth][c]{\makebox[1.2\textwidth][c]{\textsuperscript{(3)}Dublin Institute for Advanced Studies, Dublin, Ireland}}\\
  \makebox[\linewidth][c]{\makebox[1.2\textwidth][c]{\href{mailto:tianti@tcd.ie}{\textcolor{blue}{tianti@tcd.ie}}}}
}\par
\vspace{2pt}

\begin{icton_abstract}
We present a case study on the Emerald Fibre Bridge Link, an operational subsea telecom cable connecting Dublin and North Wales, examining DAS vessel-related signatures jointly with concurrent AIS data. The observations show that vessel-related DAS responses depend on local cable sensitivity and background conditions, while their interpretation is complicated by imperfect AIS reporting. Examining vessel-crossing events jointly, we identify representative DAS-AIS association patterns, ranging from clear vessel matches to offset, ambiguous, AIS-incomplete, AIS-silent-candidate, and non-vessel confounders. These observations reveal the gap between physical measurements at the cable and cooperative vessel reporting, providing practical insights for designing future DAS-assisted cable-protection workflows.
\end{icton_abstract}
\keywords{Distributed acoustic sensing; subsea telecom cables; vessel monitoring; cable protection.}

\section{INTRODUCTION}
Vessel detection using \ac{DAS} on subsea telecom cables has been demonstrated in several recent studies \cite{baird_ocean_2025,rivet_preliminary_2021,paap_leveraging_2025}. These studies show that \ac{DAS} can record vessel-radiated noise along operational cables over distances of tens to hundreds of kilometres, using array-processing, physics-based, and learning-based approaches \cite{rivet_preliminary_2021,paap_leveraging_2025,xenaki_overview_2025}. However, detection alone does not resolve how the resulting observations should be interpreted: how they should be associated with vessel identity, how uncertainty should be represented, and how monitoring systems can distinguish what is directly observed from what is inferred.

In much of the existing work, interpretation follows a common pattern: \ac{DAS} provides the detection, and \ac{AIS} provides the label \cite{rivet_preliminary_2021,paap_leveraging_2025}. A vessel signature is identified in the \ac{DAS} record, matched to an \ac{AIS} track by proximity in space and time, and the match is then used as validation \cite{rivet_preliminary_2021,baird_ocean_2025}. While this provides a practical starting point, it risks treating \ac{AIS} as ground truth and the cable as a uniform sensor, two assumptions that often break down in operational data. \ac{AIS} is a cooperative reporting system: vessels choose whether to transmit, reporting cadence varies by vessel class, and coverage near the cable can range from dense to absent \cite{emmens_promises_2021,ford_detecting_2018}. The cable, for its part, is not a uniform acoustic aperture. Background noise, sensitivity, and coupling to the seabed all vary along its length, and some of this variation is structured by the marine environment rather than by the instrumentation alone \cite{sladen_distributed_2019,mata_flores_identification_2023}.

These considerations directly affect how joint \ac{DAS}-\ac{AIS} observations are interpreted. If the cable is non-uniform, then a vessel of similar source level may be visible at one location and weak at another, while a non-vessel environmental process may trigger a detector elsewhere \cite{paap_leveraging_2025,mata_flores_identification_2023}. If \ac{AIS} support varies, the label associated with a \ac{DAS} event may be interpolated across a reporting gap, may correspond to the wrong vessel in a multi-vessel window, or may not exist at all. In those cases, the joint \ac{DAS}-\ac{AIS} observation space cannot be reduced to a simple matched/unmatched label. Understanding this structure is a necessary step before building automated vessel-detection or cable-protection systems on top of these data.

We study this problem using one week of \ac{DAS} and \ac{AIS} data from the \ac{EFBL}, an operational subsea telecom cable connecting Dublin and North Wales. From the Dublin landing station, the interrogated route comprises an initial 12 km terrestrial section followed by an approximately 120 km offshore section across the central Irish Sea. \ac{DAS} data were acquired along one fibre pair using an \ac{ASN} OptoDAS interrogator configured with a 61.28 m gauge length, 30.64 m channel spacing, and 625 Hz sampling frequency. Concurrent \ac{AIS} records at 1-min intervals provide vessel identity, position, course, and speed, allowing vessel trajectories to be projected onto the cable coordinate and \ac{CPA} to be estimated. We first examine the along-cable structure of the \ac{DAS} background and the reporting characteristics of \ac{AIS} during the observation period. We then analyze their joint behavior around vessel-crossing events. The results show that both sources are informative but non-uniform. Vessel-related \ac{DAS} signatures are therefore best interpreted by considering the local cable response together with the quality of the \ac{AIS} context.

\section{CABLE-SIDE OBSERVATIONS}

\begin{figure}[t]
  \centering
  \includegraphics[width=\textwidth]{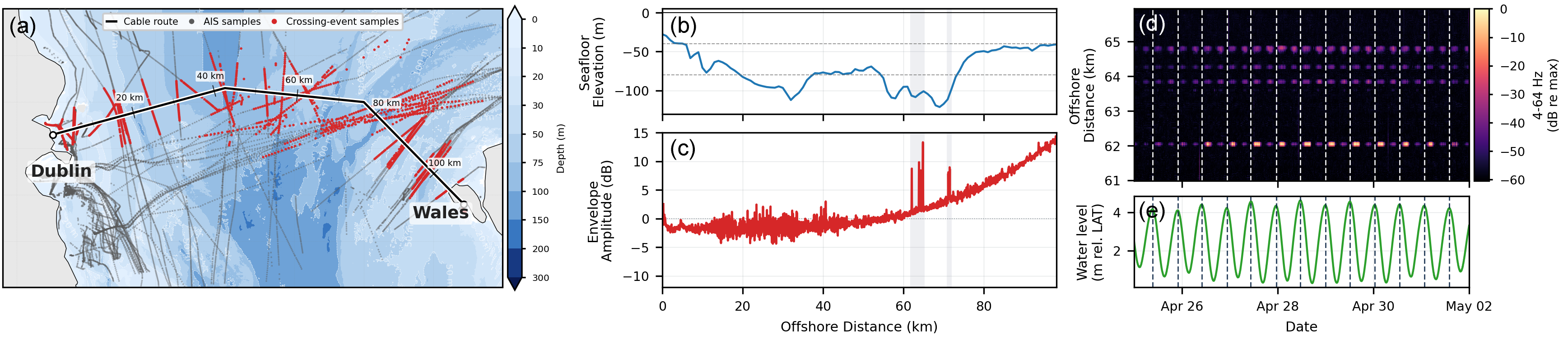}
  \caption{Experimental setting, noise analysis, and tidal modulation of anomalous channels. (a) \ac{EFBL} route, \ac{AIS} vessel coverage on 27 April 2025, and identified cable-crossing samples. (b) Bathymetric profile along the cable. (c) Per-channel median 4-64 Hz envelope over one week of data after excluding \ac{AIS}-identified vessel activity. (d) Time-distance heatmap of the 4-64 Hz envelope over the 61-66 km offshore window containing anomalous channels. (e) Water level at the Howth Water Level 1 tide gauge from the \ac{INTGN}.}
  \label{fig:study_noise_tide}
  \vspace{-5mm}
\end{figure}

Fig.~\ref{fig:study_noise_tide}(a) shows the \ac{EFBL} route across the Irish Sea together with \ac{AIS} vessel positions from 27 April 2025 and the corresponding \ac{AIS}-derived cable-crossing samples. Fig.~\ref{fig:study_noise_tide}(b) shows regional bathymetry from EMODnet \cite{emodnet_bathymetry_2024}. The cable lies in a shallow shelf-sea setting, with water depth generally increasing offshore and localized depressions reaching approximately 100-120 m.

We analyzed one week of \ac{DAS} data from 25 April to 1 May 2025. For each channel, we computed the median 4-64 Hz envelope level after excluding intervals associated with \ac{AIS}-identified vessel crossings or nearby vessel activity. The resulting profile (Fig.~\ref{fig:study_noise_tide}(c)) shows that the \ac{DAS} background is strongly non-uniform along the cable. The median level rises from 1.8 dB over the 60-70 km section to 9.8 dB over 80-100 km, approximately 11 dB higher than the 25-55 km mid-shelf section.

Several anomalous high-amplitude channels also occur at approximately 62-65 km and 71 km offshore. These features remain fixed in distance throughout the week, with levels about 12 dB and 6 dB above nearby channels respectively, but vary strongly in time. Fig.~\ref{fig:study_noise_tide}(d) shows the 4-64 Hz envelope across the 61-66 km window over the full week, revealing clear temporal structure. Comparison with sea-level observations from the Howth Water Level 1 station (53.39°N, 6.07°W) of the \ac{INTGN} \cite{marine_institute_howth_2025} (Fig.~\ref{fig:study_noise_tide}(e)) reveals a semi-diurnal modulation, with maxima near high water and reduced levels toward low water. This suggests that the response of these anomalous cable sections is tidally modulated, potentially due to differences in cable-seabed coupling, partial suspension, burial state, or other local conditions.

These observations show that the \ac{DAS} background is spatially non-uniform, structured by both marine environmental conditions and local cable state. A vessel with a given acoustic source level may produce different \acp{SNR} depending on where it interacts with the cable, and non-vessel environmental processes can dominate the signal at specific locations.

\section{AIS-SIDE OBSERVATIONS}
For \ac{DAS}-\ac{AIS} association, three properties of the \ac{AIS} stream are particularly important: the reporting cadence of individual vessels, the amount of \ac{AIS} support around each cable \ac{CPA}, and the temporal uncertainty introduced by gaps in \ac{AIS} messages near \ac{CPA}. 

\begin{figure}[t]
  \centering
  \includegraphics[width=\textwidth]{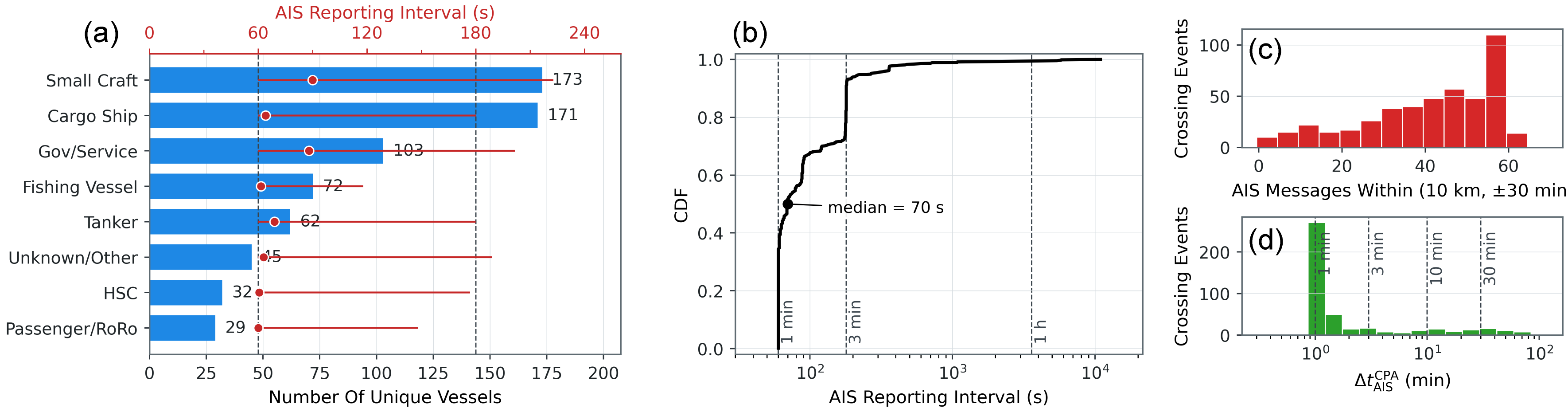}
  \caption{\ac{AIS} reporting characteristics during the seven-day analysis window.
  (a) Vessel population and typical observed reporting cadence by vessel class; dots
  show the median and whiskers show the P10-P90 range. (b) Empirical \ac{CDF} of per-vessel
  median reporting intervals. (c) Number of \ac{AIS}
  messages within 10 km of the cable and within $\pm 30$ min of \ac{CPA} across 460 cable-crossing events. (d) \ac{AIS}
  time gap across \ac{CPA} ($\Delta t_{\mathrm{AIS}}^{\mathrm{CPA}}$).}
  \label{fig:ais_quality}
  \vspace{-5mm}
\end{figure}

During the seven-day analysis window, 687 unique vessels were observed. Fig.~\ref{fig:ais_quality}(a) summarizes the vessel population and reporting cadence by class. Large commercial vessels generally report more frequently than small craft. Across classes, median reporting intervals are typically on the order of one minute, with 90th-percentile (P90) values around 3 min. The empirical \ac{CDF} of per-vessel median reporting interval (Fig.~\ref{fig:ais_quality}(b)) confirms this: the median across individual vessels is approximately 70 s, and about 91\% have a median cadence of 3 min or less.

Median cadence alone does not show whether \ac{AIS} messages are available near the moment of cable interaction. We therefore characterize support at the event level. Across the study week, 460 cable-crossing events were derived from the \ac{AIS} data. For each, we count messages within 10 km of the cable and within $\pm 30$ min of the estimated \ac{CPA}. Fig.~\ref{fig:ais_quality}(c) shows that most crossings are well supported: approximately 95\% contain at least 10 messages, with a median of 44 messages per crossing. To assess temporal uncertainty near \ac{CPA}, we compute the \ac{AIS} gap across \ac{CPA}, $\Delta t_{\mathrm{AIS}}^{\mathrm{CPA}}$, defined as the interval between the last \ac{AIS} message before and the first after the estimated \ac{CPA}. Fig.~\ref{fig:ais_quality}(d) shows that this gap is usually tight, with a median of 1.1 min. The tail, however, is non-negligible: about 5\% of crossings exceed 25 min, and the worst cases reach approximately 1.7 h. For these events, uncertainty in estimated crossing time and location is substantially increased.

Overall, the \ac{AIS} stream is sufficiently dense to provide useful context for most vessel crossings, but its quality varies across vessels and events. \ac{AIS} should therefore be treated as cooperative contextual evidence rather than exact ground truth.

\section{\texorpdfstring{JOINT \acs{DAS}-\acs{AIS} OBSERVATIONS}{JOINT DAS-AIS OBSERVATIONS}}
\begin{figure}[h]
  \centering
  \includegraphics[width=\textwidth]{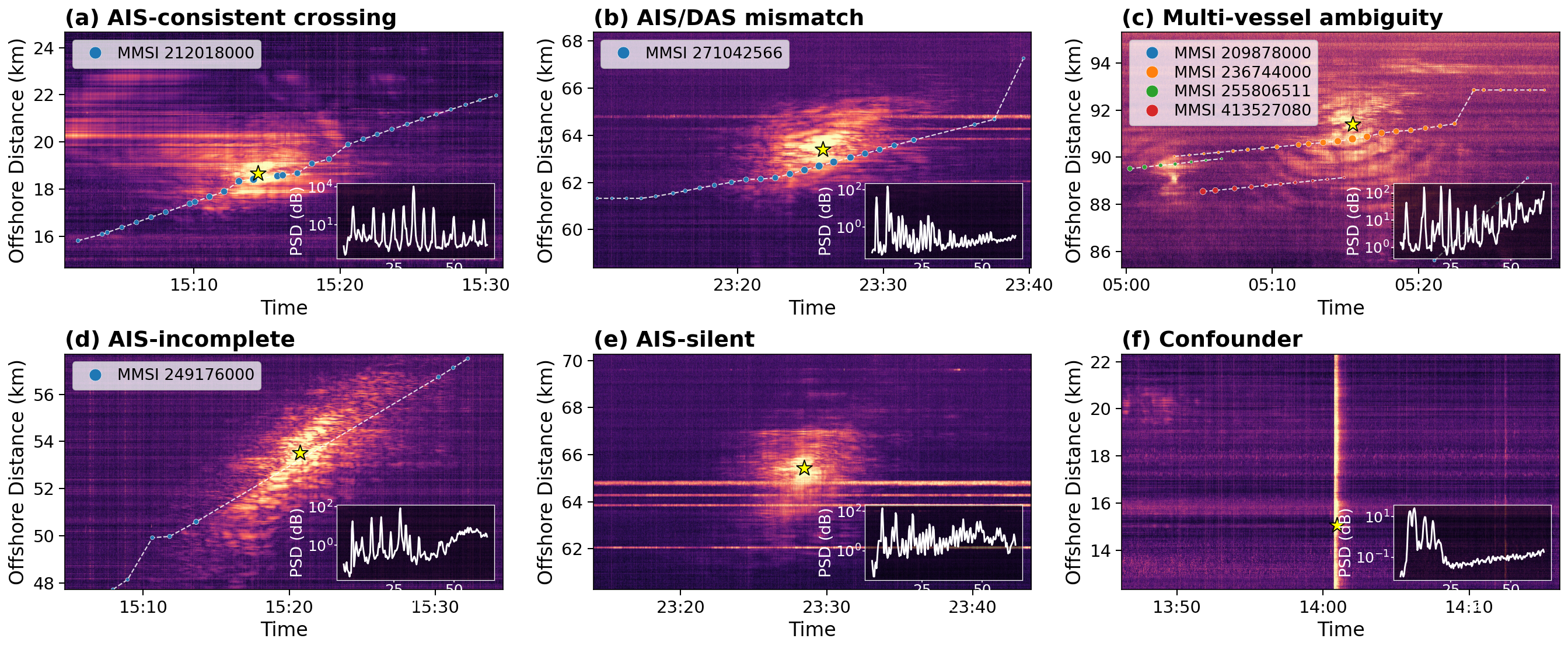}
  \caption{Representative \ac{DAS}-\ac{AIS} association patterns. Each panel shows a local time–distance window centred on a candidate event, with the 4–16 Hz \ac{DAS} envelope as the background and \ac{AIS} samples projected onto the cable coordinate as coloured markers. Dashed lines show interpolated \ac{AIS} trajectories, and the yellow star marker denotes the point used for spectral analysis.}
  \label{fig:overview_6class_seamap}
  \vspace{-5mm}
\end{figure}

Fig.~\ref{fig:overview_6class_seamap} summarizes six representative association patterns observed in the data. Each panel shows a local time–distance window centred on a candidate event, using the 4–16 Hz \ac{DAS} envelope for visualization. This band is used because most vessel-related signatures in this deployment are most clearly expressed at low frequencies. \ac{AIS} samples are projected onto the cable coordinate, and local spectral structure is shown using a \ac{PSD} extracted near the strongest \ac{DAS} response.

\textbf{\ac{AIS}-consistent crossing.} Fig.~\ref{fig:overview_6class_seamap}(a) shows the clearest association pattern. A single \ac{AIS} vessel is continuously reported through the local window, and its projected trajectory aligns with a coherent \ac{DAS} energy ridge. The local cable response is sustained in time and supports interpretation as a vessel-consistent \ac{DAS} response. 

\textbf{\ac{AIS}/\ac{DAS} mismatch.} Fig.~\ref{fig:overview_6class_seamap}(b) shows a case where a dense \ac{AIS} track and a \ac{DAS} energy ridge have similar overall structure but are offset in space and time. The \ac{AIS} trajectory lies approximately 709 m from the strongest \ac{DAS} response. This mismatch illustrates why \ac{AIS} should be treated as cooperative vessel reporting rather than a definitive reference, and why \ac{DAS}-\ac{AIS} association should retain spatial and temporal uncertainty. 

\textbf{Multi-vessel ambiguity.} Fig.~\ref{fig:overview_6class_seamap}(c) shows a window containing several \ac{AIS} vessels and multiple \ac{DAS} energy structures. Here, \ac{DAS} supports the presence of vessel-related activity near the cable, but attribution to a single vessel is uncertain. This separates two questions: whether the cable observed vessel-like activity, and which vessel produced it. In this case, choosing the correct vessel cannot be done from acoustic features alone.

\textbf{\ac{AIS}-incomplete association.} Fig.~\ref{fig:overview_6class_seamap}(d) shows a case in which a vessel track is present in the broader window, but \ac{AIS} samples are sparse and absent around the \ac{DAS} response. As a result, the trajectory and \ac{CPA} can only be estimated by interpolation over a poorly sampled interval. The \ac{DAS} ridge may be consistent with a vessel transit, but the \ac{AIS} evidence does not directly observe the vessel at the critical time and location. 

\textbf{\ac{AIS}-silent candidate.}
Fig.~\ref{fig:overview_6class_seamap}(e) shows a \ac{DAS} event with a transit-like envelope and a vessel-consistent local cable response, but no supporting \ac{AIS} samples. Such events indicate cable-observed acoustic evidence not explained by cooperative vessel reporting, thus should be treated as \ac{AIS}-silent candidates with association uncertainty.

\textbf{Confounder.} Fig.~\ref{fig:overview_6class_seamap}(f) shows a strong impulsive \ac{DAS} response that is not sufficiently vessel-like. The response is short-lived and nearly vertical in the time–distance map, lacks the pattern of a vessel transit and clear vessel-like spectral structure, and has no supporting \ac{AIS} context. This case emphasizes that high \ac{DAS} energy alone is not sufficient evidence of vessel activity; the time–distance pattern, spectral content, \ac{AIS} context, and local background must be considered together.

Together, these six cases show that \ac{DAS}-\ac{AIS} interpretation is not a one-to-one matching problem. \ac{AIS} provides cooperative contextual support; \ac{DAS} provides direct cable-observed acoustic evidence. Neither stream alone removes association uncertainty, and the recurring patterns are not independent of evidence quality on either side.  The confounder in (f) is a cable-side issue, the silent, incomplete and mismatch cases are \ac{AIS}-side issues. An operational monitoring system must therefore reason about the quality of its inputs, not just their presence.

\section{CONCLUSION}

\ac{AIS} and \ac{DAS} provide complementary observations of vessel activity: \ac{AIS} provides cooperative reports of vessel identity and motion, whereas \ac{DAS} records acoustic and vibration energy coupled into the cable. Agreement between the two is informative, but disagreement is equally important because it exposes uncertainty in vessel attribution, \ac{AIS} coverage, cable coupling, and local background conditions. We show that subsea telecom cable is not a uniform acoustic aperture. Local background structure, tidal modulation, and spatially varying sensitivity all affect whether a vessel signature is visible and how confidently it can be associated with nearby \ac{AIS} tracks. Similarly, AIS is not a ground-truth label. It provides cooperative contextual support for interpreting \ac{DAS} observations, with increased uncertainty when reports are sparse near \ac{CPA} or when multiple vessels are present.

Future automated \ac{DAS}-based vessel-monitoring systems should preserve association uncertainty, distinguish \ac{AIS}-silent candidates from cases with strong cooperative support, and include checks for local background effects and vessel-like time–distance and spectral structure. In this sense, the value of subsea cable \ac{DAS} lies not only in observing vessel-related activity, but also in providing an independent physical evidence layer for interpreting activity near critical marine infrastructure.

\small
\section*{ACKNOWLEDGEMENTS}
This publication has emanated from research conducted with the financial support of
Research Ireland and the Department of Defence under the Research Ireland-Department of
Defence grant number 24/FIP/DO/13340P, in addition to 18/RI/5721, 13/RC/2106 P2 and EU grants 101189703 (ICON) and C2024/3-3 (SUSTAINET-ADVANCE).

\end{document}